\documentclass{article}
\usepackage{graphicx}
\def\p{\partial}

\def\g{\gamma}

\def\de{\delta}

\def\De{\Delta}
\def\ov{\overline}

\def\ld{\lambda}
\def\Ld{\Lambda}

\def\ep{\epsilon}
\def\e{\eta}

\def\om{\omega}
\def\Om{\Omega}

\def\b{\beta}

\def\a{\alpha}

\def\pdellx'{\frac{\partial}{\partial x'}}
\def\pdellw'{\frac{\partial}{\partial w'}}

\newcommand{\be}{\begin{equation}}
\newcommand{\ee}{\end{equation}}
\def\bed{\begin{displaymath}}
\def\eed{\end{displaymath}}
\def\bea{\begin{eqnarray}}
\def\eea{\end{eqncrray}}
\def\[{$$}
\def\]{$$}
\begin{document}
\title{Neutrino Oscillation, Finite Self-Mass and General Yang-Mills Symmetry}

\author{ Jong-Ping Hsu\\
Department of Physics,\\
 University of Massachusetts Dartmouth \\
 North Dartmouth, MA 02747-2300, USA\\
E-mail: jhsu@umassd.edu}

\maketitle
{\small  The conservation of lepton number is assumed to be associated with a general Yang-Mills symmetry.  New transformations involve (Lorentz) vector gauge functions and characteristic phase functions, and they form a group.  General Yang-Mills fields are associated with new fourth-order equations and linear potentials.  Lepton self-masses turn out to be finite and proportional to the inverse of lepton masses, which implies that neutrinos should have non-zero masses.   Thus, general Yang-Mills symmetry could provide an understanding of neutrino oscillations and suggests that neutrinos with masses and very weak leptonic force may play a role in dark matter.}
\bigskip


\bigskip

Recently, a general Yang-Mills (gYM) symmetry was postulated to discuss implications in particle-cosmology.  It is a natural generalization of the usual gauge symmetry to deal with the intricacies of gauge fields for quark confinement and the force for late-time accelerated cosmic expansion.\cite{1,2} 
The general Yang-Mills transformations is defined by replacing space-time derivative of usual  scalar gauge functions $\p_\mu \om (x)$ and phase functions by vector gauge functions $\Ld_\mu(x)$ and characteristic phase functions respectively.  It contains the usual gauge symmetry as a special subset of vector gauge functions.  It also leads to a Lagrangian involving higher-order derivatives.\cite{3,4,5,6}  In a previous work, we used vector gauge functions and non-integrable phase functions to investigate forms of gauge fields, wrapping number and quantization conditions in  gauge field theories.\cite{7}  Similarly, we further discussed vector gauge functions and gYM symmetry, which led to a new form of gauge field associated with fourth-order equations and static linear potentials.  The linear potential produced by color charges of general Yang-Mills symmetry with $SU_3$ group could provide an explicit confinement mechanism for quarks.\cite{8}  Moreover, the corresponding r-independent force produced by extremely small point baryon-charges associated with $U_1$ could provide an explanation for the late-time accelerated cosmic expansion.\cite{1}

We employed gYM symmetry to explore further physical implications of fourth-order gauge field equations in the lepton sector.
It was speculated and hoped that Lagrangians and Hamiltonians involving higher order derivatives would help physicists to formulate a finite quantum field theory.\cite{3,4,5}  In this paper, we formulated a quantum leptonic dynamics with a general leptonic $U_1$ (i.e., $U_{1\ell}$) symmetry associated with the conservation of the lepton number (or charge).  The gYM transformations have group properties and its symmetry leads to a new general $U_{1\ell}$ field.  We obtain the rules for Feynman diagrams and calculate the lepton self-mass at the one-loop level.  We employ the dimensional regularization, which is convenient to preserve the gauge invariance of the S matrix.   We show that the self-mass of leptons, including neutrinos, is finite and proportional to the inverse of the lepton mass.   Their implications will be discussed below.

The new Yang-Mills symmetry generalizes the usual phase function $\om(x)$ in $exp[-i g \om (x)] $ to a characteristic phase function $P(x)$,\cite{9}
\be
exp[-ig \om (x)]  \ \   \to  \ \   exp[-ig P(x)]\equiv  exp\left(-ig \int_{x'_o}^{x'_e=x} d x'^\ld \Ld_\ld (x')\right)_{Le},
\ee
where we have a fixed point $x'_o$ and a variable end point $x'_e=x$.
The subscript `$Le$' in (1) denotes that the path in the integration is required to satisfy the `Lagrange equation'
\be
 [\p_\mu \Ld_\ld(x)-\p_\ld \Ld_\mu(x)]dx^\mu=0, \ \ \ \ \   c=\hbar=1.
\ee
This type of  additional condition for the path will make the gYM transformations in previous works\cite{1,2,8} unambiguous.  Thus, we have a well-defined property for the characteristic phase function $P(x)$,
\be
\p_\mu  P(x) = \Ld_\mu(x).
\ee
Such a new phase function $P(x)$ in (1) and (3) resembles Hamilton's characteristic function, and is well-defined.\cite{10,11}

In leptonic dynamics, the gYM transformations for the general $U_{1\ell}$ fields $L_\mu(x)$ and the lepton $\ell(x)$ are defined by
$$
L'_\mu(x) = L_\mu(x) + \Ld_\mu(x),  \ \ \  
$$
\be
 \ell'(x) = \Om(x) \ell(x), \ \ \  \ov{\ell}'(x) = \ov{\ell}(x) \Om^{-1}(x),
\ee
$$
\Om(x)= exp(-iP(x)).
$$
The vector gauge functions $\Ld_\mu(x)$ are assumed to satisfy the constraint equations,
\be
\p^\mu\p_\mu \Ld_\nu (x) - \p^\mu\p_\nu \Ld_\mu (x) = 0.
\ee
 The solutions of the second-order partial differential equations (5) form a set of infinitely many functions.  For the gYM transformations (4) to be definable, the characteristic phase factor $\Om(x)$ in (4) must be single valued.  Nevertheless, the characteristic phase function $P(x)$ need not be.  

For each vector gauge function $\Ld_\mu^a(x)$, there are compatible equations (2) and (5) with $\Ld_\mu(x)=\Ld_\mu^a(x)$, so that the relation (3), i.e., $\p_\mu  P^a(x) = \Ld^a_\mu(x)$ holds.\cite{9,10,11}  The characteristic phase function so defined satisfies the group property, e.g.
\be
\Om^c(x)=\Om^a(x) \Om^b(x),   \  \ or  \ \  \Ld^c(x)=\Ld^a(x) + \Ld^b(x),
\ee
where $\Ld^c(x)$ satisfies
$$
\p^\mu\p_\mu \Ld^c_\nu (x) - \p^\mu\p_\nu \Ld^c_\mu (x) = 0, \ \ \   [\p_\mu \Ld^c_\ld(x)-\p_\ld \Ld^c_\mu(x)]dx^\mu=0.
$$

As usual, the gauge-covariant derivative ${\De}_{\mu} $ in leptonic dynamics is defined by,  
\be
\De_\mu =\p_\mu +ig L_\mu(x).
\ee
Morevoer, the gauge curvature $L_{\mu\nu}(x) $ associated with the general $U_{1\ell}$ symmetry can be obtained from the commutator of  $\De_{\mu}$. We have  
\be
[\De_{\mu} ,\De_{\nu}]=   ig L_{\mu\nu}(x), 
\ee
$$
L_{\mu\nu}(x) = \p_\mu L_\nu(x) - \p_\nu L_\mu(x).
$$
 We stress that the gauge curvature $L_{\mu\nu}(x)$ is not invariant under the general Yang-Mills transformations (4) involving the vector gauge function $\Ld_\mu(x)$ because 
\be
  L'_{\mu\nu}(x)= L_{\mu\nu}(x) +\p_\mu \Ld_\nu (x) - \p_\nu \Ld_\mu (x) \ne L_{\mu\nu}(x),
\ee
where $\Ld_\mu(x) \ne \p_\mu \om(x)$.  However, the space-time derivative of the gauge curvature, i.e., $\p^\mu L_{\mu\nu}$, is invariant under the new gauge transformations, 
\be
\p^\mu L'_{\mu\nu}(x)= \p^\mu L_{\mu\nu}(x) +\p^\mu\p_\mu \Ld_\nu (x) - \p^\mu\p_\nu \Ld_\mu (x)  =\p^\mu L_{\mu\nu}(x),
\ee
where we have used the constraint (5).  We also have
\be
\ov{\ell}' \g^\mu \De'_{\mu} \ell' (x) =  \ov{\ell} \g^\mu \De_{\mu} \ell (x).
\ee
Thus, the gYM invariant Lagrangian is assumed to be quadratic in the `invariant curvature' $\p^\mu L_{\mu\nu}$,
\be
L_{lep}= \frac{L^2_{s}}{2}(\p^\mu L_{\mu\ld}) \p_{\nu}L^{\nu \ld}
 + i\ov{\ell} \g^\mu (\p_\mu  + ig L_\mu)\ell - m_\ell \ov{\ell} \ell.
\ee
where the summation in the terms involving leptons, $ \ell=(e, \nu_e, \mu, \nu_\mu, \tau, \nu_\tau)$,  is understood.  

One may wonder what is the relation between the new $U_{1\ell}$ transformations in (4)-(5) and the usual $U_1$ gauge transformations?  We note that although there are four components of arbitrary gauge vector function $\Ld_\mu(x)$ in (4), they are required to satisfy four constraints in the form of partial differential equations (5).   In the special case when the vector function $\Ld_\mu(x)$ can be expressed as the space-time derivative of an arbitrary (usual) scalar function $\om(x)$, i.e., $\Ld_{\mu}=\p_{\mu} \om(x)$, the constraint equations (2) and (5) become identities for arbitrary function $\om(x)$ and the transformations (4)-(5) become the same as the usual $U_1$ gauge transformations with arbitrary scalar function $\om(x)$.  In other words, the characteristic phase function $P(x)$ in (1) become usual phase function of the $U_1$ group.  Thus, the new transformations (4) together with the constraint (5) may be considered as  generalized $U_1$ gauge transformations.   Their corresponding gauge fields  $L_\mu(x)$ may be called general $U_1$ (gauge) fields.   The quanta of the quantized lepton gauge fields  $L_\mu(x)$ may be called `leon', which satisfy the fourth-order equations derived from (12), as shown in (19) below.

As usual, the complete Feynman-Dyson rules for Feynman diagrams can be derived from the total Lagrangian with the gauge fixing terms $L_{gf}$, 
\be
L_{tot}=L_{\ell} + L_{gf},  \ \ \  L_{gf} =  \frac{\xi}{2}(\p^{\ld}\p_{\mu}L^{\mu})(\p_{\ld}\p_{\nu}L^{\nu}),    
\ee
The propagator of leons can be obtained from their  fourth-order field equations, which can be derived from the  Lagrangian (12).  The leon propagator is    
\be
\ell_{\mu\nu}(k)=\frac {-i}{L^2_s (k^2+i\ep)^2}\left[\e_{\mu\nu} - \left(1-\frac{1}{\xi}\right)\frac{k_\mu k_\nu}{k^2+i\ep}\right].
\ee
The  rules for the  leon-lepton 3-vertex   
[$\overline{{\ell}}(k_1) \ell(k_2) L_{\mu}(k_3)$]  is given respectively by
\be
     -i g \g_{\mu}.   
 \ee
Other rules such as fermion propagators, a factor -1 for each fermion loop etc. are the same as those in usual QED.\cite{12}  

To see new physical effects of the fourth-order field equations dictated by the general $U_{1\ell}$ symmetry let us consider the lepton self-mass $\de m$.  Following the Feynman-Dyson rules for writing the invariant amplitude $-i M_{fi}$, we have
\be
-i \de m=(-ig)^2 \int \frac{d^4 k}{(2\pi)^4}  \ell^{\mu\nu}(k) \frac{i \g_\mu [\g \cdot (p-k) + m]\g_\nu}{(p-k)^2 - m^2 + i\ep},
\ee
where the leon propagator $\ell^{\mu\nu}(k)$ is given by (14), which has a better high energy behavior than the photon propagator.
Using dimensional regularization (see appendix), we obtain
\be
\de m = \frac{3 g^2}{16 \pi^2 L^2_s m}.
\ee
Thus, the lepton self-mass is finite in quantum leptonic dynamics based on the invariant Lagrangian (12).  In comparison with the electron self-mass $\de m_e$ in QED, its divergence is usually expressed in terms of the ultraviolet  cut-off $\Ld$,\cite{12} 
\be
\de m_e = \frac{3e^2 m_e}{16 \pi^2} \ ln \left(\frac{\Ld^2}{m_e^2}\right),   
\ee
which corresponds to the term involving $1/(4-D)$ in the dimensional regularization. We note that quantum leptonic dynamics is not a completely finite theory\footnote{In a sense, the quantum leptonic dynamics with gYM symmetry may be termed a `super-renormalizable' theory} because the leon self-energy is still divergent, just like the photon self-energy  in QED.\cite{12}  Nevertheless, this problem is related to gauge symmetry and can be handled by the dimensional regularization.

 Experimentally, the neutrino mass has an upper limit of 2 eV.  One may get a rough estimate of the unknown parameter by a reasonable conjecture, $\de m/ m < 1 $.  One has the ratio $g^2/L^2_s < 1/(10^{-15} m^2)$. Furthermore, there is one interesting qualitative property in (17) for any neutrinos, namely, $\de m \propto 1/m$.  This property implies that all neutrinos should have non-zero masses, which make neutrino oscillations possible, in consistent with recent neutrino experiments.   In this connection, we note that the electroweak theory does not incorporate neutrino masses naturally.  Perhaps, the result (17) could be considered as an experimental support of the general Yang-Mills symmetry associated with the conservation law of lepton charge. 

In the past decades, it has been considered that the $U_1$ corresponding to lepton number is not a local gauge symmetry because the massless gauge field coupled to lepton number has not been observed.  We take the view that this may simply imply that the lepton charge $g$ is extremely small.

As we have mentioned previously, the invariant Lagrangian (12) leads to a fourth-order equation for leptonic gauge field $L_\mu(x)$
\be
\p^2 \p^\ld L_{\ld\mu} - \frac{g}{L_s^2}\ov{\psi} \g_\mu \psi = 0.
\ee
For the static case with a point lepton-charge $g$ at the origin, (19) lead to a linear potential $V(r)$,
\be
\nabla^2 \nabla^2 L_0({\bf r})=\frac{g}{L^2_s}\de^2({\bf r}), \ \ \ \  V(r)=gL_0({\bf r})=\frac{-g^2 r}{8\pi L^2_s},
\ee
which leads to repulsive and r-independent force between two point lepton-charge of the same sign.  
However, when one considers the force between, say, a point lepton-charge and a uniform sphere of lepton-charges, the resultant force could be modified.\cite{13}

Baryon charges also have similar properties, as discussed in previous works.\cite{1,13}  Suppose these forces are much weaker than the gravitational force, so that they cannot be detected in the solar system or in the Milky Way galaxy.  Nevertheless, they may be able to give an explanation of the late-time accelerated cosmic expansion.    The reason is as follows:  After the universe expands for a long time,  when the distance between two baryon/lepton galaxies are large enough, then the r-independent baryon-lepton repulsive force will overcome the gravitational force and causes the late-time accelerated cosmic expansion.\footnote{In this case, the dark matter will move along with the baryon/lepton galaxies, provided the dark matter particles are bonded to baryon/lepton by the stronger gravitational force.}
Furthermore, the three known stable neutrinos are the final products of almost all decay processes of known particles\cite{14} that might exist in the beginning of the universe.  We expect that there are enormous amount of neutrinos in the late-time universe after particle-antiparticle annihilations and particle decays, even though they may be negligible when the universe was `created' because they do not have strong and electromagnetic interactions. It is possible that these enormous amounts of stable neutrinos with small masses and a very weak leptonic force may help us understand the mystery of dark matter and, hence, deserve to be further investigated.

The fourth-order field equation is usually considered as unphysical because the dynamical system involves non-definite energy.\cite{3,6}  However, the leon cannot be directly detected because of its extremely small coupling constant.  On the other hand, suppose one considers quark model with gYM symmetry,\cite{2,8} the gauge bosons with large coupling constants are permanently confined in the quark system and, hence, cannot be detected as a free particle with negative energies. Nevertheless, other new ideas are needed for a satisfactory S matrix with the general Yang-Mills symmetry. For example, the model with gYM symmetry also suggests that these new gauge bosons with fourth-order equation could be consistently treated as off-mass-shell particles, so that they do not appear in external physical states of the S matrix and do not contribute imaginary part of amplitudes when they appear in internal states of a process. These properties related to unitarity will be discussed in a separate paper.

In light of previous discussions, the general Yang-Mills symmetry appears to be interesting and useful because, apart from giving various linear potentials for quark confinement and for late-time accelerated cosmic expansion\cite{1,2,13,15}, it may also be able to provide a field-theoretic basis for understanding finite lepton self-mass, neutrino masses and neutrino oscillations. 

The work was supported in part by the Jin-Shin Research Fund of the UMassD Foundation.

\bigskip 

\noindent
{\small { \bf Appendix}  \ \ \  Finite Lepton Self-Mass}

Let us calculate the lepton self-mass (16) by using D-dimensional regu-larization\cite{16,17} and replacing  $k^2$ by $k^2+\ld^2$ in the leon propagator (14) to define mathematics involving possible infrared divergence.  We shall take the limits $\ld \to 0$ and $D \to 4$ at the end of calculations.   We write (16) in the following form
\be
-i \de m=\frac{-g^2}{L_s^2}\left[T_1 - \left(1-\frac{1}{\xi}\right)T_2 \right].
\ee
Let us consider $T_1$ first,
\be
T_1=\int^1_0 2xdx \int  \frac{d^D k}{(2\pi)^D}  \frac{ (2-D)\g_\mu (p-k)^\mu + Dm}{[k^2 -2k_\nu p^\nu (1-x) +(p^2 - m^2)(1-x)+ \ld^2 x]^3},
\ee
where the $i\ep$ prescription in the propagator is understood.  After the shift of variable $k_\mu \to k_\mu + p_\mu(1-x)$ and carrying out the dimensional regularizations of the self-mass, we obtain
\be
T_1=\frac{-im}{8\pi^2}(2\a - \b), 
\ee
where it is safe to set $D=4$ and then calculate $\a$ and $\b$.  We obtain 
\be
\a=\int ^1_0 \frac{xdx}{R}=\left[\frac{1}{2A} ln|R| - \frac{B Z}{2A} \right]^1_0= \frac{1}{2m^2} \left(ln\frac{\ld^2}{m^2} - 1\right), 
\ee
\be
\b=\int ^1_0 \frac{x^2dx}{R}=\left[\frac{x}{A}-\frac{B}{2A^2} ln|R| + \frac{Y Z}{2A^2} \right]^1_0=\frac{1}{m^2} \left(\frac{1}{2}+ ln\frac{\ld^2}{m^2}\right),
\ee
\be
R=Ax^2 +Bx + C, \ \ A =p^2, \ \ \ B= -( p^2+m^2+\ld^2), \ \ \ C= m^2, 
\ee
\be
Y=B^2-2AC, \ \ \   Z=\frac{1}{\sqrt{B^2 -4AC}} ln\frac{|2Ax+B-\sqrt{B^2-4AC} \ |}{|2Ax+B +\sqrt{B^2 -4AC}\ |}.
\ee
Thus, $T_1$ is given by
\be
T_1= \frac{3 g^2}{16 \pi^2 L^2_s m}.
\ee
Let us calculate the gauge-dependent term $T_2$ in (21).  We have
\be
T_2=  \int  \frac{d^D k}{(2\pi)^D}  \frac{\g_\mu k^\mu \g_\nu p^\nu\g_\ld k^\ld -  \g_\mu k^\mu \g_\nu k^\nu\g_\ld  k^\ld+ m\g_\mu k^\mu \g_\nu k^\nu }{(k^2 +\ld^2)^3 [(p-k)^2   - m^2]},
\ee
$$
=\int  \frac{d^D k}{(2\pi)^D}  \left[\frac{\g_\mu k^\mu(p^2-m^2) - (\g_\mu p^\mu - m)k^2 }{(k^2 +\ld^2)^3[(p^2 - 2p^\mu k_\mu  - m^2]} -\frac{\g_\mu k^\mu}{(k^2 + \ld^2)^3}\right] = 0,
$$
where we have used  $\g \cdot p = m$ and $p^2 = m^2$ because $\de m$ is sandwiched between free lepton spinors.   We also use the relation $\int d^D k \ f(k^2) k^\mu$ = 0 in the dimensional regularization.

 It follows from (21), (28) and (29) that the lepton self-mass is finite, as shown in (17).  It is interesting that the two infrared logarithmic-divergent terms in (24) and (25) cancel each other, as shown in (23).  We also note that it is important to use the small parameter $\ld^2 > 0$ to define the mathematics related to possible infrared divergence.\footnote{This appears to be consistent with the Fourier sine transform of the generalized function ${1/(\bf k}^2)^2$, which is related to the linear potential.\cite{1,18}}  Suppose  one uses $\ld^2 < 0$ in the calculations of the integrals $\a$ and $\b$ in (24) and (25).  One will have an arctangent function rather than the logarithmic function $Z$ in (27) and one will have two infrared divergent terms $\propto 1/\ld$ in $\a$ and $\b$, which do not cancel each other.

\bibliographystyle{unsrt}


\end{document}